\documentclass[aps,preprint]{revtex4}
\usepackage{amssymb,epsf}
\usepackage{amsmath}
\usepackage{graphicx}
\begin{document}
\title{Topological Black Holes of (n+1)-dimensional Einstein-Yang-Mills Gravity}
\author{N. Bostani$^1$\footnote{email address:
nbostani@ihep.ac.cn} and M. H. Dehghani$^{2,3}$\footnote{email address:
mhd@shirazu.ac.ir}}
\affiliation{$^1$ Key Laboratory of Particle Astrophysics, Institute of High Energy Physics,
Chinese Academy of Sciences, Beijing 100049, China\\
$^2$Physics Department and Biruni Observatory, College of Sciences, Shiraz University, Shiraz 71454, Iran\\
$^3$ Research Institute for Astrophysics and Astronomy of Maragha (RIAAM), Maragha,
Iran}

\begin{abstract}
We present the topological solutions of Einstein gravity in the presence
of a non-Abelian Yang-Mills field. In ($n+1$) dimensions, we consider the
$So(n(n-1)/2-1,1)$ semisimple group as the Yang-Mills gauge group, and
introduce the black hole solutions with hyperbolic horizon. We argue
that the 4-dimensional solution is exactly
the same as the 4-dimensional solution of Einstein-Maxwell gravity, while the higher-dimensional
solutions are new. We investigate the
properties of the higher-dimensional solutions and find that these solutions in 5 dimensions have the same properties as the
topological 5-dimensional
solution of Einstein-Maxwell (EM) theory although the metric function
in 5 dimensions is different. But in 6 and higher dimensions,
the topological solutions of EYM and EM gravities with non-negative mass have different properties.
First, the singularity of EYM solution does not present a naked singularity
and is spacelike, while the singularity of topological Reissner-Nordstrom
solution is timelike. Second,
there are no extreme 6 or higher-dimensional black holes in EYM gravity with non-negative mass,
while these kinds of solutions exist in EM gravity. Furthermore, EYM theory has no
static asymptotically de Sitter solution with non-negative mass, while EM gravity has.
\end{abstract}

\maketitle

\section{Introduction}

The Einstein-Yang-Mills (EYM) theory, which explains the theory of a
gravitating non-Abelian gauge field, may be regarded as the most natural
generalization of Einstein-Maxwell gravity. The solitonic solution of this
theory in 4 dimensions has been discovered by Bartnik and McKinnon for the
gauge group $SU(2)$ \cite{BK}. The colored black hole solutions of this
theory with $SO(3)$ gauge group have been introduced in \cite{Yas}, while
those with $SU(2)$ gauge group have been investigated in \cite{YMblack1,YMblack2,YMblack3}.
These solutions have led to certain revisions of some of the basic concepts
of black hole physics based on the uniqueness and no-hair theorem. It is now
well-known that this theory possesses ''hairy'' black hole solutions, whose
metric is not a member of the Kerr-Newmann family (see \cite{Volkov} for a
detailed review in 4 dimensions and \cite{Radu1} for a recent review in
higher dimensions). Furthermore, unlike the Kerr-Newmann black holes, the
geometry exterior to the event horizon is not determined uniquely by global
charges measurable at infinity, although only a small number of parameters
are required in order to describe the metric and matter field \cite{Eliza}.
Solutions of the EYM equations in higher dimensions have been also studied
in \cite{Oku,Halil}. These solutions were also extended in the presence of
cosmological constant \cite{Maeda,Vol2,Mann,Hosotani,Eliz1} and Gauss-Bonnet term \cite
{Tch1,Tch2}.

From stability analysis, it turns out that the solution with zero or
positive cosmological constant is unstable \cite{Brod}, while those with
negative cosmological constant are stable \cite{Hosotani,Eliza2}. The
presence of a negative cosmological constant also invites the topological
black holes into the game. Indeed, the horizon topology of an asymptotically
flat black hole should be a round sphere, while in AdS space it is possible
to have a black holes with zero or negative constant curvature horizon too.
These black holes are referred to as topological black holes in the
literature, and investigated by many authors \cite{Top1,Top2,Top3,Top4,Top5,Top6,Top7}. All of these
investigations are mainly based on Einstein-Maxwell theory, however
numerical solutions have been considered in \cite{Radu2,Radu3,Ma2} in the presence of $%
SU(2)$ Yang-Mills field. It may be of interest to generalize these
topological solutions for a non-Abelian matter field and investigate their
properties. Indeed, the analysis of Einstein's equation with nonlinear field
sources may shed new light on the generic properties of topological
solutions of Einstein's equations. In this paper, we want to study the ($n+1$%
)-dimensional topological black hole solutions of EYM gravity with a
negative cosmological constant.

The outline of the paper is as follows. We give a brief review of the field
equations of EYM gravity for a semisimple gauge group in Sec. \ref{Fiel}. In
Sec. \ref{4d} we first present the 4-dimensional solution for gauge group $%
SO(2,1)$ and investigate its properties, and second we introduce the
5-dimensional solution which incorporates a logarithmic term unprecedented
in other dimensions. By a similar analogy we extend these solutions in $%
(n+1) $ dimensions with gauge group $So(n(n-1)/2-1,1)$. We finish our paper
with some concluding remarks.

\section{Field equations\label{Fiel}}

The model which will be discussed here is an ($n+1$)-dimensional EYM system
for an $N$-parameters gauge group ${\cal G}$, which is assumed to be at
least semisimple with structure constants $C_{bc}^{a}$. The metric tensor of
the gauge group is
\[
\Gamma _{ab}=C_{ad}^{c}C_{bc}^{d},
\]
where the Latin indices $a$, $b$.... go from $1$ to $N$, and the repeated
indices is understood to be summed over. According to Cartan's criteria the
determinant of $\Gamma _{ab}$ is not zero, and therefore one may define
\[
\gamma _{ab}\equiv -\frac{\Gamma _{ab}}{\left| \det \Gamma _{ab}\right|
^{1/N}},
\]
where $\left| \det \Gamma _{ab}\right| $ is the positive value of
determinant of $\Gamma _{ab}$. The action of $(n+1)$-dimensional EYM gravity
with negative cosmological constant $\Lambda =-n(n-1)/2l^{2}$ may be written
as
\begin{equation}
I_{{\rm EYM}}=\int d^{n+1}x\sqrt{-g}[R+\frac{n(n-1)}{l^{2}}-\gamma
_{ab}F_{\mu \nu }^{(a)}F^{(b)\mu \nu }],  \label{Act1}
\end{equation}
where $R$ is the Ricci Scalar and $F_{\mu \nu }^{(a)}$ is the gauge field
tensor defined as:
\begin{equation}
F_{\mu \nu }^{\left( a\right) }=\partial _{\mu }A_{\nu }^{\left( a\right)
}-\partial _{\nu }A_{\mu }^{\left( a\right) }+\frac{1}{2e}C_{bc}^{a}A_{\mu
}^{\left( b\right) }A_{\nu }^{\left( c\right) }.  \label{gfield}
\end{equation}
In Eq. (\ref{gfield}) $e$ is a coupling constant and $A_{\mu }^{\left(
a\right) }$'s are the gauge potentials. Variation of the action (\ref{Act1})
with respect to the spacetime metric $g_{\mu \nu }$ and the gauge potential $%
A_{\mu }^{(a)}$ yield the EYM equations as
\begin{eqnarray}
&& F^{(a)\mu \nu }_{\ \ \ \ \ ;\nu} =j^{(a)\mu},  \label{FE2} \\
&& G_{\mu \nu }+\Lambda g_{\mu \nu } =8 \pi T_{\mu \nu },  \label{FE1}
\end{eqnarray}
where the gauge current and the stress-energy tensor carried by the gauge
fields are
\begin{eqnarray}
&& j^{(a)\nu}=\frac{1}{e}C_{bc}^{a}A_{\mu }^{(b) }F^{(c)\mu \nu},
\label{cur} \\
&& T_{\mu \nu }=\frac{1}{4\pi}\gamma _{ab}\left( F_{\mu }^{(a)\lambda
}F_{\nu \lambda }^{(b)}-\frac{1}{4}F^{(a)\lambda \sigma }F_{\lambda \sigma
}^{(b)}g_{\mu \nu }\right),  \label{EMt}
\end{eqnarray}
respectively.

\section{Topological Black Holes In 4 and 5 Dimensions\label{4d}}

The 4-dimensional static metric of a topological spacetime with a hyperbolic
horizon may be written as
\begin{equation}
ds^{2}=-f(r)dt^{2}+\frac{dr^{2}}{f(r)}+r^{2}\left( d\theta ^{2}+\sinh
^{2}\theta d\varphi ^{2}\right) .  \label{Metr1}
\end{equation}
Introducing the coordinates
\begin{eqnarray*}
x_{1} &=&r\sinh \theta \cos \varphi , \\
x_{2} &=&r\sinh \theta \sin \varphi , \\
x_{3} &=&r\cosh \theta ,
\end{eqnarray*}
and using the Wu-Yang ansatz \cite{Wu}, one obtains the explicit form of YM
potentials as
\begin{eqnarray}
A^{(1)} &=&\frac{e}{r^{2}}\left( x_{1}dx_{3}-x_{3}dx_{1}\right) =-e\left(
\cos \varphi d\theta -\sinh \theta \cosh \theta \sin \varphi d\varphi
\right) ,  \nonumber \\
A^{(2)} &=&\frac{e}{r^{2}}\left( x_{2}dx_{3}-x_{3}dx_{2}\right) =-e\left(
\sin \varphi d\theta +\sinh \theta \cosh \theta \cos \varphi d\varphi
\right) ,  \nonumber \\
A^{(3)} &=&\frac{e}{r^{2}}\left( x_{1}dx_{2}-x_{2}dx_{1}\right) =e\sinh
^{2}\theta d\varphi ,  \label{YM4}
\end{eqnarray}
which have the Lie algebra of $SO(2,1)$ with nonzero structure constants and
$\gamma _{ab}$ as follows:
\begin{eqnarray*}
C_{23}^{1} &=&C_{31}^{2}=-C_{12}^{3}=1, \\
\gamma _{ab} &=&{\rm diag}(-1,-1,1).
\end{eqnarray*}
Now, it is a matter of calculation to show that the YM fields (\ref{YM4})
satisfy the YM field equation (\ref{FE2}), while the gauge currents (\ref
{cur}) don't vanish and are
\begin{eqnarray}
&&j^{(1)}=\frac{e}{r^{4}}(\cos \varphi d\theta+\coth \theta \sin \varphi
d\varphi) ,  \nonumber \\
&&j^{(2)}=-\frac{e}{r^{4}}(\sin \varphi d\theta-\coth \theta \cos \varphi
d\varphi) ,  \nonumber \\
&&j^{(3)}=-\frac{e}{r^{4}}d\varphi ,  \label{J4}
\end{eqnarray}
Here, it is worth to mention that one may perform a position-dependent gauge
transformation from the gauge field (\ref{YM4}) to a set of three Abelian
gauge fields which satisfy the Maxwell equation with zero current
independently \cite{Yas}. Of course the scalar
\begin{equation}
{\cal F}\equiv \gamma _{ab}F^{(a)\mu \nu }F_{\mu \nu }^{(b)},  \label{F2}
\end{equation}
is invariant under this transformation. In $4$ dimensions, ${\cal F}%
=2e^{2}/r^{4}$ for both the solutions of YM and Maxwell equations, and
therefore the solution of EYM is the same as the topological solution of
Reissner-Nordstrom solution.

To find the function $f(r)$, one may use any components of Eq. (\ref{FE1}).
The simplest equation is the $rr$ component of these equations which can be
written as
\begin{equation}
\left[ r(1+f)\right] ^{\prime }=\frac{3}{l^{2}}r^{2}-\frac{e^{2}}{r^{2}},
\label{Eqf4}
\end{equation}
where prime denotes the derivative with respect to $r$. The solution of Eq. (%
\ref{Eqf4}) is
\begin{equation}
f(r)=-1+\frac{r^{2}}{l^{2}}-\frac{2m}{r}+\frac{e^{2}}{r^{2}},  \label{F4}
\end{equation}
where $m$ is an integration constant which is related to the mass of the
spacetime. Of course the above solution satisfies all the other components
of the field equations. This solution has the same properties as the
asymptotically AdS topological solution of EM gravity in 4 dimensions, and
we do not discuss it more here. Also it is worth to mention that this
solution is the counterpart of the spherical solution of EYM theory
introduced in \cite{Yas}.

The 5-dimensional solution incorporates a logarithmic term unprecedented in
other dimensions, and therefore we shall treat it in some details.
Recently, static non-abelian black hole solutions of five-dimensional
maximal (${\cal N} = 8$) gauged supergravity has been considered in Ref. \cite{Cvetic}.
Here, we consider the
5-dimensional, static metric with hyperbolic horizon which may be written as
\begin{equation}
ds^{2}=-f(r)\;dt^{2}+\frac{dr^{2}}{f(r)}+r^{2}\left\{ d\theta ^{2}+\sinh
^{2}\theta \left( d\varphi ^{2}+\sin ^{2}\varphi d\psi ^{2}\right) \right\} .
\label{Met2}
\end{equation}
Introducing the coordinates
\begin{eqnarray}
x_{1} &=&r\sinh \theta \sin \varphi \cos \psi ,  \nonumber \\
x_{2} &=&r\sinh \theta \sin \varphi \sin \psi ,  \nonumber \\
x_{3} &=&r\sinh \theta \cos \varphi ,  \nonumber \\
x_{4} &=&r\cosh \theta ,  \nonumber
\end{eqnarray}
and using the ansatz
\begin{eqnarray}
A^{(1)} &=&\frac{e}{r^{2}}\left( x_{1}dx_{4}-x_{4}dx_{1}\right) ,  \nonumber
\\
A^{(2)} &=&\frac{e}{r^{2}}\left( x_{2}dx_{4}-x_{4}dx_{2}\right) ,  \nonumber
\\
A^{(3)} &=&\frac{e}{r^{2}}\left( x_{3}dx_{4}-x_{4}dx_{3}\right) ,  \nonumber
\\
A^{(4)} &=&\frac{e}{r^{2}}\left( x_{1}dx_{2}-x_{2}dx_{1}\right) ,  \nonumber
\\
A^{(5)} &=&\frac{e}{r^{2}}\left( x_{1}dx_{3}-x_{3}dx_{1}\right) ,  \nonumber
\\
A^{(6)} &=&\frac{e}{r^{2}}\left( x_{2}dx_{3}-x_{3}dx_{2}\right) ,  \nonumber
\\
\end{eqnarray}
one obtains:
\begin{eqnarray}
A^{(1)} &=&-e\sin \varphi \cos \psi d\theta -e\sinh \theta \cosh \theta
(\cos \varphi \cos \psi d\varphi -\sin \varphi \sin \psi d\psi ),  \nonumber
\\
A^{(2)} &=&-e\sin \varphi \sin \psi d\theta -e\sinh \theta \cosh \theta
\left( \cos \varphi \sin \psi d\varphi +\sin \varphi \cos \psi d\psi \right)
,  \nonumber \\
A^{(3)} &=&-e\cos \varphi d\theta +e\sinh \theta \cosh \theta \sin \varphi
d\varphi ,  \nonumber \\
A^{(4)} &=&e\sinh ^{2}\theta \sin ^{2}\varphi d\psi  \nonumber \\
A^{(5)} &=&-e\sinh ^{2}\theta \left( \cos \psi d\varphi -\sin \varphi \cos
\varphi \sin \psi d\psi \right) ,  \nonumber \\
A^{(6)} &=&-e\sinh ^{2}\theta \left( \sin \psi d\varphi +\sin \varphi \cos
\varphi \cos \psi d\psi \right) .  \label{YM5}
\end{eqnarray}
The non-zero structure constants of the group are:
\begin{eqnarray}
C_{24}^{1} &=&C_{35}^{1}=C_{41}^{2}=C_{36}^{2}=C_{51}^{3}=C_{62}^{3}=1,
\nonumber \\
C_{56}^{4} &=&C_{21}^{4}=C_{64}^{5}=C_{31}^{5}=C_{45}^{6}=C_{32}^{6}=1,
\end{eqnarray}
which show that the gauge group is isomorphic to $So(3,1)$. Again, it ia a
matter of calculations to show that the gauge currents are
\begin{eqnarray}
&&j^{(1)}=\frac{2e}{r^{4}}\left( \sin \varphi \cos \psi d\theta +\coth
\theta \cos \varphi \cos \psi d\varphi -\frac{\coth \theta }{\sin \varphi }%
\sin \psi d\psi \right) ,  \nonumber \\
&&j^{(2)}=\frac{2e}{r^{4}}\left( \sin \varphi \sin \psi d\theta +\coth
\theta \cos \varphi \sin \psi d\varphi -\frac{\coth \theta }{\sin \varphi }%
\cos \psi d\psi \right) ,  \nonumber \\
&&j^{(3)}=-\frac{2e}{r^{4}}\left( \cos \varphi d\theta +\coth \theta \sin
\varphi d\varphi \right) ,  \nonumber \\
&& j^{(4)} =-\frac{2e}{r^{4}}d\psi ,  \nonumber \\
&& j^{(5)}=\frac{2e}{r^{4}}\left( \cos \psi d\varphi -\cot \varphi \sin \psi
d\psi \right) ,  \nonumber \\
&& j^{(6)}=\frac{2e}{r^{4}}(\sin \psi d\varphi +\cot \varphi \cos \psi d\psi
).  \label{J5}
\end{eqnarray}
Calculating the left hand-side of YM equation (\ref{FE2}), one obtains
exactly the same expressions as (\ref{J5}). Thus the gauge fields (\ref{YM5}%
) satisfy the YM field equation (\ref{FE2}). Using the definition of the
metric tensor of the gauge group, one obtains:
\[
\gamma _{ab}={\rm diag}(-1,-1,-1,1,1,1),
\]
The value of the invariant ${\cal F}$ for the YM fields (\ref{YM5}) is
\begin{equation}
{\cal F}_{{\rm YM}}=\frac{6e^{2}}{r^{4}},  \label{F2YM}
\end{equation}
while for spherically symmetric solutions of Maxwell theory is
\begin{equation}
{\cal F}_{{\rm Max}}=\frac{2e^{2}}{r^{6}}.  \label{F2Max}
\end{equation}
Comparison of the invariant ${\cal F}$ for YM and Maxwell fields given in
Eqs. (\ref{F2YM}) and (\ref{F2Max}) shows that one cannot introduce a
position-dependent transformation from the non-Abelian gauge fields to a set
of Abelian ones which satisfy the Maxwell equation. This guarantees that the
Yasskin theorem which states that the solutions of EYM and EM theories are
the same \cite{Yas} does not hold in 5 dimensions. Also, it is worth to
mention that the $r$-dependence of the components of energy momentum tensor
for EYM and EM theories are not the same. This point has been discussed in
details in the next section.

The $rr$ component of the field equation (\ref{FE1}) reduces to
\[
\left[ r^{2}(1+f)\right] ^{\prime }=\frac{4}{l^{2}}r^{3}-\frac{2e^{2}}{r},
\]
with the solution
\begin{equation}
f(r)=-1-\frac{3m}{r^{2}}-\frac{2e^{2}\ln (r)}{r^{2}}+\frac{r^{2}}{l^{2}},
\label{F5}
\end{equation}
where $m$ is an integration constant which is related to the mass of the
spacetime. Of course, the metric function (\ref{F5}) satisfies all the other
components of the EYM field equations.

\begin{figure}
\centering {\includegraphics[width=7cm]{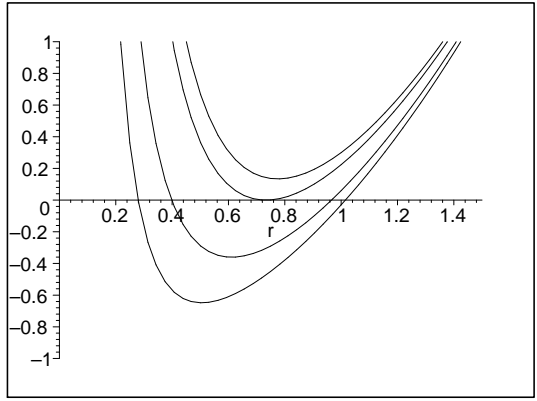} }
\caption{$f(r)$ versus $r$ for $n=4$, $l=1$, $e=.2$,
$m<m_{\mathrm{ext}}<0$, $m=m_{\mathrm{ext}}<0$, $m_{\mathrm{ext}}<m<0$
and $m>0$ from up to down, respectively.} \label{Fr5}
\end{figure}

In order to study the general structure of the solution given in Eq. (\ref
{F5}), we first look for curvature singularities. It is easy to show that
the Kretschmann scalar $R_{\mu \nu \lambda \kappa }R^{\mu \nu \lambda \kappa
}$ diverges at $r=0$ and so the above metric given by Eqs. (\ref{Met2}) and (%
\ref{F5}) has an essential singularity at $r=0$. Since the function $f(r)$
goes to $\infty $ as $r$ goes to zero and becomes $+\infty $ as $r$ goes to
infinity, the singularity is timelike. Seeking possible black hole
solutions, we turn to look for the existence of horizons. The horizon(s) is
(are) located at the roots of $g^{rr}=f(r)=0$. The function $f(r)$ may have
zero, one or two real roots. Denoting the largest real root of $f(r)$ by $%
r_{+}$, we consider first the case that $f(r)$ has one real root. In this
case $f^{\prime }(r)$ vanishes at
\begin{equation}
r_{{\rm ext}}=\frac{l}{2}\left(1+ \sqrt{1+\frac{8e^{2}}{l^{2}}}\right).
\label{rc4} \\
\end{equation}
The value of mass for which the metric function has one real root may be
obtained as
\begin{equation}
m_{{\rm ext}}=-\left\{\frac{e^{2}}{3}\left[ \ln \left( \frac{l^{2}+l\sqrt{%
l^{2}+8e^{2}}}{4}\right) -\frac{1}{2}\right] +\frac{l^{2}+l\sqrt{l^{2}+8e^{2}%
}}{24}\right\}.  \label{mex5}
\end{equation}
Then, the metric of Eqs. (\ref{Met2}) and (\ref{F5}) presents a naked
singularity provided $m<m_{{\rm ext}}$, an extreme black hole for $m=m_{{\rm %
ext}}$ and a black hole with two horizons if $m>m_{{\rm ext}}$. Figure \ref
{Fr5} shows the diagram of $f(r)$ for various values of the mass parameter.

One may note that although the metric function (\ref{Fr5}) has a logarithmic
term which is absent in the 5-dimensional Reissner-Nordstrom solution, the
properties of these two solutions are the same. The Hawking temperature of
the black holes can be easily obtained by requiring the absence of conical
singularity at the horizon in the Euclidean sector of the black hole
solutions. One obtains
\begin{equation}
{T}_{+}=\frac{2r_{+}^{4}-l^{2}r_{+}^{2}-e^{2}l^{2}}{2\pi l^{2}r_{+}^{3}},
\label{Temp5}
\end{equation}
which vanishes for the extreme black hole with $m=m_{{\rm ext}}$ given in
Eq. (\ref{mex5}).

\section{Higher dimensional Solutions\label{nd}}

We assume that the metric has the following form:
\begin{equation}
ds^{2}=-f(r)\;dt^{2}+\frac{dr^{2}}{f(r)}+r^{2}\left( d\theta ^{2}+\sinh
^{2}\theta d\;\Omega _{n-2}^{2}\right) ,  \label{Metn}
\end{equation}
where $d\Omega _{n-2}^{2}$ is the line element of $(n-2)$-sphere. In order
to obtain the gauge fields, we use the coordinates
\begin{eqnarray*}
x_{1} &=&r\sinh \theta \prod_{j=1}^{n-2}\sin \varphi _{j}, \\
x_{i} &=&r\sinh \theta \cos \varphi _{n-i}\prod_{j=1}^{n-i-1}\sin \varphi
_{j};\text{ \ \ }i=2...n-1, \\
x_{n} &=&r\cosh \theta ,
\end{eqnarray*}
and the ansatz
\begin{eqnarray}
A^{(a)} &=&\frac{e}{r^{2}}\left( x_{i}dx_{n}-x_{n}dx_{i}\right) ;\text{ \ \
\ \ }a=i=1...n-1,  \nonumber \\
A^{(b)} &=&\frac{e}{r^{2}}\left( x_{i}dx_{j}-x_{j}dx_{i}\right) ;\text{ \ \
\ \ \ \ }i<j,  \label{YMn}
\end{eqnarray}
where $b$ runs from $n$ to $n(n-1)/2$. It is a matter of calculation to show
that the Lie algebra of the gauge group is $So(n(n-1)/2-1,1)$ with the
following $\gamma _{ab}$:
\[
\gamma _{ab}=\epsilon _{a}\delta _{ab};\text{ \ \ no sum on }a,
\]
where $\epsilon _{a}$ is
\begin{equation}
\epsilon _{a}=\left\{
\begin{array}{ll}
-1 & \ \ \ 1\leq a\leq n-1 \\
1 & \ \ \ n\leq a\leq \frac{n(n-1)}{2}
\end{array}
\right.   \label{kappa}
\end{equation}
One can also shows that the above gauge fields (\ref{YMn}) satisfy the YM
field equation (\ref{FE2}), while the gauge currents do not vanish.

Again, one may calculate the value of the invariant ${\cal F}$ for the $%
n(n-1)/2 $ non-Abelian gauge fields (\ref{YMn}) and $n(n-1)/2$ Abelian gauge
fields of Maxwell equation with spherical symmetry as
\begin{eqnarray}
{\cal F}_{{\rm YM}} &=&\frac{(n-1)(n-2)e^{2}}{r^{4}}, \\
{\cal F}_{{\rm Max}} &=&\frac{2e^{2}}{r^{2(n-1)}}.
\end{eqnarray}
Thus, one cannot introduce a transformation from the non-Abelian gauge
fields to a set of Abelian ones which satisfy the Maxwell equation. This
shows that the solutions of EYM theory are not the same as EM theory in $%
(n+1)$ dimensions with $n>3$. Also, one may show that the $r$-dependence of
the components of energy momentum tensor for EYM and EM theories are not the
same. Using the definition of the energy-momentum tensor (\ref{EMt}), one
obtains:
\begin{eqnarray}
&&T_{\phantom{t}{t}}^{t}=T_{\phantom{r}{r}}^{r}=-\frac{(n-2)(n-1)e^{2}}{%
2r^{4}},  \nonumber \\
&&T_{\phantom{\theta}{\theta}}^{\theta }=T_{\phantom{\varphi_i}{\varphi_i}%
}^{\varphi_i}=-\frac{(n-2)(n-5)e^{2}}{2r^{4}},  \label{EMtYM}
\end{eqnarray}
while the energy-momentum of Maxwell field may be written as:
\begin{eqnarray}
&&\left(T^t_{\phantom{t}{t}}\right)_{{\rm Max}}=\left(T^r_{\phantom{r}{r}%
}\right)_{{\rm Max}}\sim -\frac{e^{2}}{r^{2(n-1)}},  \nonumber \\
&&\left(T^{\theta}_{\phantom{\theta}{\theta}}\right)_{{\rm Max}%
}=\left(T^{\varphi_i}_{\phantom{\varphi_i}{\varphi_i}}\right)_{{\rm Max}%
}\sim \frac{e^{2}}{r^{2(n-1)}}.  \label{EMtM}
\end{eqnarray}
As one may note by comparing Eqs. (\ref{EMtYM}) and (\ref{EMtM}), the $r$%
-dependence of energy-momentum of these two fields are different for $n>3$.
Also the tangential components of these two energy momentum tensor differ by
a minus sign for $n>5$ and is zero for $n=5$. These differences show that
the Yasskin theorem which states that the solution of EYM and EM theories
are the same \cite{Yas} is only true in 4 dimensions. Also, as we see below
these differences drastically change the properties of the solutions.

The $rr$-component of the field equation (\ref{FE1}) reduces to
\begin{equation}
\left[ r^{n-2}(1+f)\right] ^{\prime }=\frac{n}{l^{2}}%
r^{n-1}-(n-2)e^{2}r^{n-5}=0.  \label{Eqfn}
\end{equation}
Integrating Eq. (\ref{Eqfn}), one obtains
\begin{equation}
f(r)=-1+\frac{r^{2}}{l^{2}}-\frac{(n-1)m}{r^{n-2}}-\frac{(n-2)e^{2}}{%
(n-4)r^{2}};\text{ \ \ }n\neq 4.  \label{Fn}
\end{equation}
\begin{figure}
\centering {\includegraphics[width=7cm]{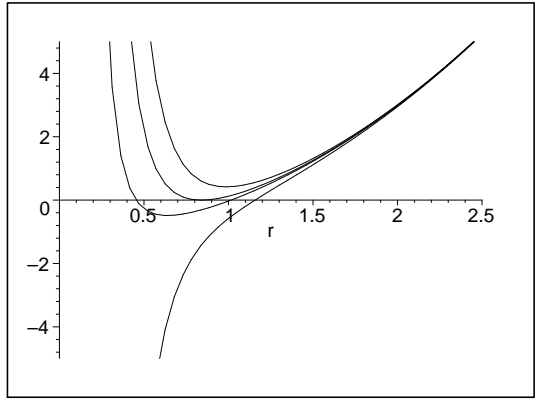} }
\caption{$f(r)$ versus $r$ for $n=6$, $l=1$, $e=.2$,
$m<m_{\mathrm{ext}}$, $m=m_{\mathrm{ext}}$,
$m_{\mathrm{ext}}<m<0$ and $m>0$ from up to down, respectively.} \label{Fr6}
\end{figure}
Unlike the topological Reissner-Nordstrom solutions in 6 and higher
dimensions, the singularity at $r=0$ for the solutions with non-negative
mass is spacelike, and therefore it is unavoidable. This is due to the fact
that $f(r)$ approaches to $-\infty $ as $r$ goes to zero and goes to $%
+\infty $ at large $r$. These solutions present black holes with one
horizon. The spacetime of Eqs. (\ref{Metn}) and (\ref{Fn}) with negative
mass presents a naked singularity if $m<m_{{\rm ext}}<0$, an extreme black
hole for $m=m_{{\rm ext}}<0$ and a black hole with inner and outer horizons
provided $m>m_{+{\rm ext}}$, where $m_{{\rm ext}}$ is
\begin{eqnarray}
m_{{\rm ext}} &=&-2\frac{r_{{\rm ext}}^{n-4}}{n-1}\left( \frac{r_{{\rm ext}%
}^{4}}{(n-2)l^{2}}+\frac{e^{2}}{n-4}\right) ,  \nonumber \\
r_{{\rm ext}} &=&\sqrt{\frac{n-2}{2n}}l\left\{ 1+\sqrt{1+\frac{4ne^{2}}{%
(n-2)l^{2}}}\right\} ^{1/2}.  \label{rexn}
\end{eqnarray}
Figure \ref{Fr6} shows $f(r)$ in term of $r$ for various values of $m$.

The metric of the extreme black hole near horizon may be written as
\[
ds^{2}=-C_{0}(r-r_{{\rm ext}})^{2}\;dt^{2}+\frac{dr^{2}}{C_{0}(r-r_{{\rm ext}%
})^{2}}+r^{2}\left( d\theta ^{2}+\sinh ^{2}\theta d\;\Omega
_{n-2}^{2}\right) ,
\]
where
\[
C_{0}=\frac{f^{^{\prime \prime }}(r_{{\rm ext}})}{2}=\frac{4n}{l^{2}}\frac{1+%
\frac{e^{2}}{(n-2)^{2}l^{2}}+\sqrt{1+\frac{4ne^{2}}{(n-2)l^{2}}}}{\left( 1+%
\sqrt{1+\frac{4ne^{2}}{(n-2)l^{2}}}\right) ^{2}}.
\]
Thus, the structure of the spacetime near horizon is almost the same as that
of  Einstein-Maxwell gravity. The only difference is that the horizon radius
of these extreme black holes can be obtained analytically [see Eq. (\ref
{rexn})], and are smaller than those of Einstein-Maxwell theory. Also the
value of $C_{0}$ is different.

\subsection{Thermodynamical properties}

The Hawking temperature is given by
\[
T_{+}=\frac{nr_{+}^{4}-(n-2)l^{2}r_{+}^{2}-(n-2)e^{2}l^{2}}{4\pi
l^{2}r_{+}^{3}},
\]
which vanishes for $m=m_{{\rm ext}}$. The entropy is given by
\[
S=\frac{V_{n-1}}{4}r_{+}^{n-1},
\]
where $V_{n-1}$ is the volume of the constant $t$ and $r$ hypersurface with
radius 1. Now we want to compute the mass of the system. One may note that
the ADM mass diverges as in the case of solutions introduced in Ref. \cite
{Tch3}. Using the first law of thermodynamics, one may obtain the
thermodynamical mass (see Ref. \cite{Oku} for more details) through the use
of the relation $T_{+}=\left( \partial M_{T}/\partial S\right) _{e}$ as
\[
M_{T}=\frac{(n-1)^{2}V_{n-1}}{16\pi }m.
\]
The above equation shows that the parameter $m$ may be denoted as the mass parameter as
mentioned before. It is worth to mention that these results
are valid in all dimensions.

\section{Closing Remarks}

In this paper, we introduced the topological black holes of
Einstein-Yang-Mills theory with hyperbolic horizon and investigate their
properties. The topological solution of EYM gravity in 4 dimensions is not a
new solution, and is exactly the same as the topological solution of EM
theory. This is due to the fact that one can perform a position-dependent
gauge transformation, from the $So(2,1)$ gauge fields of YM theory and a set
of three Abelian gauge fields of Maxwell theory. But, we showed that one
cannot perform a position-dependent gauge transformation from the $%
So(n(n-1)/2-1,1)$ gauge fields of YM theory and a set of Abelian gauge
fields of Maxwell equation in 5 and higher dimensions. Also, we noted that
the components of energy-momentum tensor of YM fields are proportional to $%
r^{-4}$ which are drastically different from the components of
energy-momentum of a Maxwell field which are proportional to $r^{-2(n-1)}$.
Indeed, not only the $r$-dependence of the components of energy-momentum of
the $So(n(n-1)/2-1,1)$ YM fields ($n>3$) is different from the
energy-momentum components of a $U(1)$ gauge field, but also they are
different in a minus sign which has a drastic effect on the properties of
the solutions [see the minus sign in front of the last term in the metric
function (\ref{Fn})]. That is, the EYM solutions in 5 and higher dimensions
are new topological solutions.

These solutions in higher dimensions with negative mass have the same
properties as the solutions in EM gravity. Here, it is worth to compare the
distinguishing features of non-negative mass solutions of EYM and EM
gravities in 6 and higher dimensions. First, these solutions of EYM theory
do not present a naked singularity, while those of EM gravity do. That is,
the solutions of EYM theory respect the cosmic censorship hypothesis.
Second, the singularity in the case of EYM black hole is spacelike and
therefore unavoidable, while the singularity of EM black holes are timelike.
Third, the solutions in EYM gravity cannot present an extreme black hole,
while those of EM gravity have extreme black hole solutions. Furthermore,
unlike the Reissner Nordstrom black holes, the geometry exterior to the
event horizon is not determined by a global charge measurable at infinity.

Also, it is worth to mention that one does not have a static asymptotically
de Sitter solution in 6 and higher-dimensional EYM gravity with non-negative
mass, while in EM theory this kind of solution exists. This is due to the
fact that the metric function of higher-dimensional Einstein-Yang-Mills
gravity in the presence of a positive cosmological constant, $%
\Lambda=n(n-1)/l^2$, may be written as
\[
f(r)=-1-\frac{r^{2}}{l^{2}}-\frac{(n-1)m}{r^{n-2}}-\frac{(n-2)e^{2}}{%
(n-4)r^{2}};\text{ \ \ }n\geq 5
\]
which is negative everywhere for $m\geq0$, and therefore the solution is not
static.

{\bf Acknowledgements}

This work has been supported by Research Institute for Astrophysics and
Astronomy of Maragha. Neda Bostani acknowledges funding support from the National Natural Science Foundation of China
under grant Nos. 10725313 and 10821061, and also likes to thank S. H. Hendi for his comments
on this work.

\end{document}